\documentclass[
preprint,
5p,
times,
twocolumn,
]{elsarticle}

\usepackage{graphicx}
\usepackage{dcolumn}
\usepackage{bm}
\usepackage{units}
\usepackage{url}
\usepackage[mathlines]{lineno}
\usepackage{varwidth,xcolor}
\usepackage{subfigure}
\usepackage{color}
\usepackage{float}
\usepackage{natbib}

\begin{document}

\title{A Novel Electrical Method to Measure Wire Tensions for {\color{black} Time Projection Chambers} }
\author[UoM]{Diego Garcia-Gamez\corref{cor1}}
\ead{diego.garciagamez@manchester.ac.uk}
\author[UoM]{Vincent Basque}
\author[UoM]{Thomas G. Brooks\fnref{shef}}
\author[UoM]{Justin J. Evans}
\author[UoM]{Michael Perry}
\author[UoM]{Stefan S{\"o}ldner-Rembold}
\author[UoM]{Fabio Spagliardi\fnref{oxf}}
\author[UoM]{Andrzej M. Szelc}

\cortext[cor1]{Corresponding Author}
\fntext[shef]{now at the University of Sheffield}
\fntext[oxf]{now at the University of Oxford}
\address[UoM]{School of Physics and Astronomy, The University of Manchester, Oxford Road, Manchester, M13 9PL, United Kingdom}

\begin{abstract}
We present a novel electrical technique to measure the tension of wires in multi-wire drift chambers. We create alternating electric fields by biasing adjacent wires on both sides of a test wire with a superposition of positive and negative DC voltages on an AC signal ($V_{\rm AC} \pm V_{\rm DC}$). The resulting oscillations of the wire will display a resonance 
at its natural frequency, and the corresponding change of the capacitance will lead to a measurable current. This scheme is scalable to multiple wires and therefore enables us to precisely measure the tension of a large number of wires in a short time. This technique can also be applied at cryogenic temperatures making it an attractive solution for future large time-projection chambers such as the DUNE detector. We present the concept, an example implementation and its performance in a real-world scenario and discuss the limitations of the sensitivity of the system in terms of voltage and wire length.
\end{abstract}
\begin{keyword}
LArTPC; APA; Wire tension; Electrical
\end{keyword}

\maketitle
\section{Introduction}
The next generation of neutrino experiments will use Liquid-Argon Time-Projection Chambers (LArTPCs) as detector technology~\cite{Rubbia:1977zz}. In these detectors, the ionization electrons produced in the passage of a charged particle drift from their production points towards planes of anode wires, called Anode Plane Assemblies (APAs). The wires are strung or wrapped at different orientations along the length of a metallic frame. A minimum of two planes is required for three-dimensional reconstruction of the events in the liquid argon. The pitch between wire planes and between wires in a plane is typically a few millimeters (3--5\,mm), which defines the spatial resolution of the LArTPC. With the increasing size of these detectors, ranging from a few hundred tons for the short-baseline neutrino (SBN) experiments~\cite{Antonello:2015lea} to 10\,kt for a single DUNE module~\cite{Acciarri:2016ooe}, the numbers of wires can go from tens of thousands to hundreds of thousands. 

When these detectors cool down to liquid-argon temperatures, the thermal contraction in the thin wires can occur faster than in the steel frame. For that reason, the nominal tension requirement chosen for the wires must be small enough to prevent breaking of wires during cool down, while being strong enough to ensure even the longest wires do not sag, as this could introduce readout noise~\cite{Acciarri:2017sde}. To prevent wire breakage or sagging, the completed wire planes need to be surveyed to ensure the wire tension is within specifications. Currently, this is the most time consuming step in the production of APAs. The standard method used to measure the tension employs a laser-based optical setup that measures the fundamental frequency of the wire, which is related to the wire tension $T$ through

\begin{equation} \label{eq:tension}
T = 4 \mu L^{2} f_0^{2}   \; ,
\end{equation}

where $\mu$ is the wire's linear mass density, $L$ its length, and $f_0$ its fundamental frequency in Hz. To perform the laser measurement, the wires have to be mechanically excited by, for example, strumming~\cite{Acciarri:2016ugk} or compressed air~\cite{Baldini:2007iba,Wang:2017mbk}. The usage of this mechanical method in large-scale detector production is difficult to envision due to how time consuming it can become and the difficulty of replicating the automated positioning system at the final assembly site. To mitigate sagging and breakage, wires are split into segments by using holders or combs, similar to capodasters, at intervals along the wire length. This can result in different tensions in each of the segments, requiring separate measurements for each segment, thus further complicating the procedure.

Different electromagnetic-based techniques of measuring wire tensions in wire chambers have been described in the literature~\cite{Ohama:1997ww,Hoshi:1985,Brinkley:1996xb,Lang:1998yw,Andryakov:1998vk}. Here, we present a novel method using an electric-based system which resolves most of the difficulties of the laser-based setup. This technique has the ability to measure multiple wires simultaneously, does not require a mechanical disturbance of the wires and permits measurements of tension at {\color{black} cryogenic} temperature, even when the cryostat of the detector is closed. 
{\color{black} In this work we have demonstrated our method in a stable environment of cooled down gaseous nitrogen, at a temperature comparable to that of liquid argon.} This last feature has never been available in LArTPC detectors and could be used to monitor the tension change during cooling in situ and identify any problems arising from the cooling process, providing the possibility to correct them immediately.

\begin{figure}[htbp]
  \centering
\includegraphics[width=0.3\textwidth]{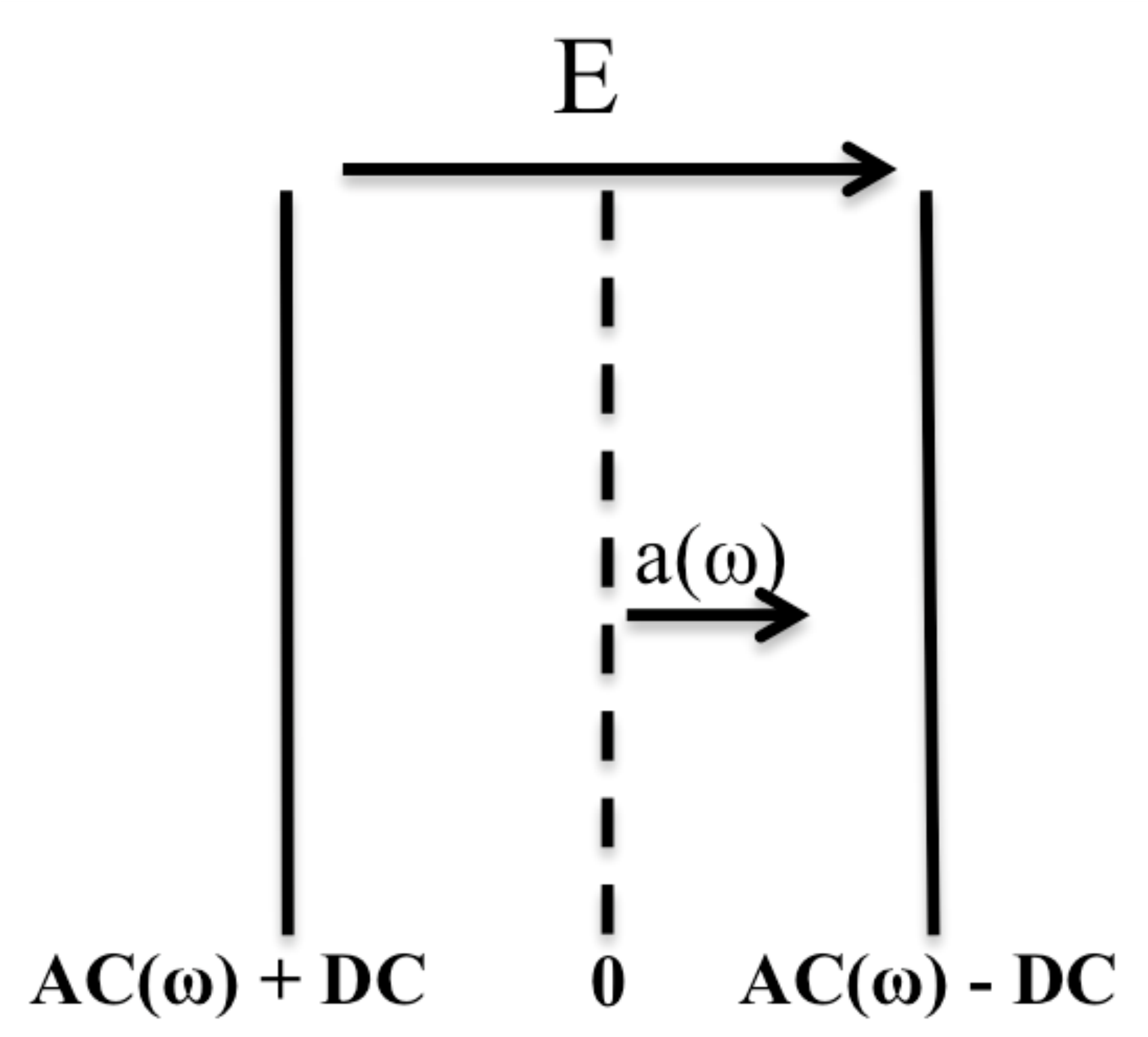}
  \caption{Schematic of the electrical tension measurement system. The ground wire is denoted by a dashed line; it is the wire under measurement. Wires denoted with a solid line are biased with a combination of AC and DC voltages, the latter with a changing sign. This results in an enhanced electric force on the wire under test.} \label{fig:cartoon}
\end{figure}

\section{Description of the system} \label{sec:description}

Each pair of wires in a multi-wire chamber acts as a capacitor. As proposed in Ref.~\cite{Ohama:1997ww}  an electric force is created
by applying AC and DC electric fields, which causes oscillating movements of the wires. The displacements of the wires change the capacitance of the two-wire system.
When the AC frequency corresponds to the natural frequency of either wire the system encounters a resonance, making the displacement particularly large and the change in capacitance measurable through the change in current flowing between the wires. If the driving force has the form $F\propto \sin(\omega t)$, the wires will move as a driven damped harmonic oscillator. 

In the setup discussed here we employ a system of three wires (see Figure~\ref{fig:cartoon}). The wire under test is kept at ground potential, while the two adjacent wires in a wire plane are biased by an AC voltage with added or subtracted DC voltages ($V_{\rm AC} \pm V_{\rm DC}$) on either side. At the resonance frequency, the wires will oscillate as in a two-wire system. For the ground wires, the opposite electric fields on each side will enhance the resultant electric force responsible for the movement. This novelty carries with it two major improvements: (i) it breaks the ambiguity introduced by the technique using two wires \cite{Ohama:1997ww} and (ii) it requires relatively low voltages, of the order of a few hundred volts. 

The need for this low bias voltage is driven by the design of modern TPC detectors where wires are usually held by termination boards built to apply the plane bias voltage, which is on the order of few hundred volts. Since in normal TPC operation the wires are never exposed to large potential differences, the electric components and PCB tracks on these boards are typically not designed to work at high voltages, limiting the maximum bias voltage that can be applied between adjacent wires in a plane. Previously proposed electrical tension measurement methods~\cite{Ohama:1997ww} require placing a large bias voltage, as large as 1.7\,kV, across the wires to obtain a reasonable signal. Using such a method in LArTPC detectors would thus require installing appropriately rated resistors and capacitors to accommodate this voltage, which would be prohibitively expensive. In addition, the wire pitch itself might also limit the maximum space between PCB tracks making it even harder to apply the needed high voltage bias. Our method's need for significantly smaller bias voltages across each component makes it immediately applicable to modern LArTPC detectors. 

\begin{figure}
\centering
\includegraphics[width=0.45\textwidth]{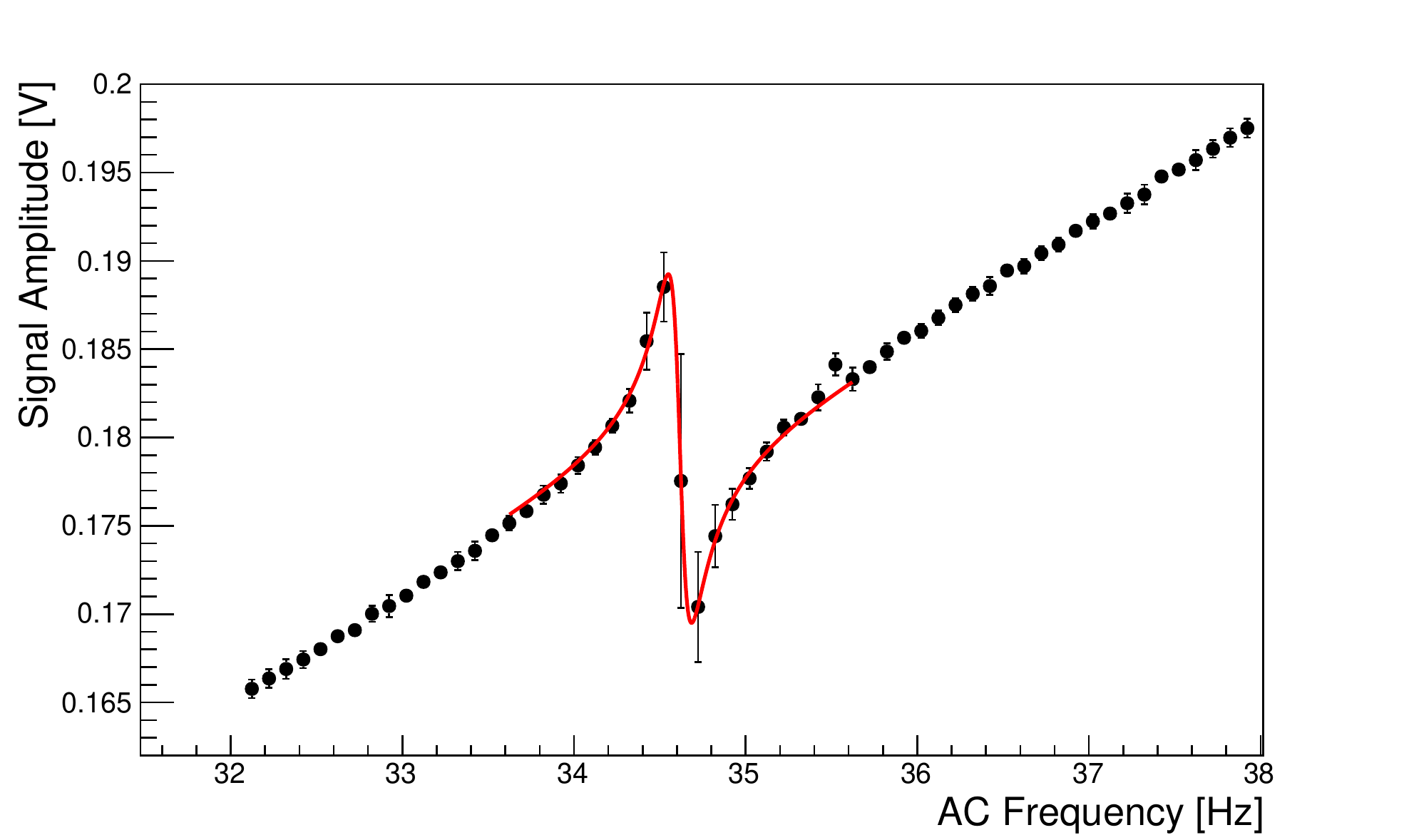}
\caption{Frequency scan of a wire with the bipolar resonance peak of the data points fitted with the model described by Equation~\ref{eq:model} (red line).}
\label{fig:model}
\end{figure}

For simplicity, we use a two wire model in order to describe the signals that will be obtained. If the wires are in air (or gas nitrogen as in Section~\ref{sec:cold}), we can assume that the system is underdamped, and the amplitude $a(\omega)$ of the driven oscillation for a given angular frequency $\omega$ of the driving force will factorize as
\begin{equation}
a(\omega) = a_e(\omega) + a_{\rm const} \; ,
\label{eq:amp}
\end{equation} 
where
\begin{equation}\label{eq:ae} 
a_e(\omega)\propto V_{\rm AC}V_{\rm DC}\frac{\omega_{0}^{2} - \omega^{2}}{(\omega_{0}^{2} -
\omega^{2})^2 + (2\Gamma\omega)^2}=V_{\rm AC}V_{\rm DC}f(\omega)
\end{equation}
and 
\begin{equation}
a_{\rm const}\propto\frac{V_{\rm DC}^{2}}{\omega_{0}^{2}} \; ,
\end{equation} 
where $\Gamma$ is the damping coefficient of the system and $\omega_{0}=2\pi f_0$ the natural angular frequency of the wire. We have folded the resonant behaviour into $f(\omega)$. The expected current read out from the circuit, assuming $a(\omega)$ $\ll$ wire pitch, can then be described as
\begin{eqnarray}
\lefteqn{I(t) = c_1V_{\rm AC}\omega\cos(\omega t) - c_2V_{\rm DC}\omega a_e(\omega)\cos(\omega t) +\nonumber}\\
& & +~{\mathcal O}(V_{\rm AC}a_e(\omega))  \nonumber\\
& & \approx \left[c_1V_{\rm AC}\omega - c_2V_{\rm DC}^{2}V_{\rm AC} \omega f(\omega)\right]\cos(\omega t) \; ,
\label{eq:model}
\end{eqnarray}
with constants $c_1$ and $c_2$ that depend on the system. Equation~\ref{eq:model} shows that if $V_{\rm AC} \ll V_{\rm DC}$, then the amplitude of the output current in the frequency domain will have the form shown in Figure~\ref{fig:model}. The term proportional to $c_1$  gives the underlying linear rise in amplitude, proportional to the driving frequency. The term proportional to $c_2$ introduces a bipolar resonance that centers on the natural frequency of the wire, and scales with the voltages as $V_{\rm AC} \times V_{\rm DC}^2$. The amplitude of the signal at resonance also scales with $1/(4\Gamma(\omega_0 - \Gamma))$ in the maximum and $-1/(4\Gamma(\omega_0 + \Gamma))$ in the minimum. This means that for a given tension $n \times T_0$, the amplitude ratio changes as $1/\sqrt{n}$. In Section~\ref{sec:voltage} we compare the behaviour of a system composed of three wires to this prediction. In a realistic system, additional effects may play a role, such as the presence of the other wire planes which could also affect the amplitude. These will be tested in a future work. 

A tension measurement becomes a scan over frequencies of the driving AC voltage, and a search for the bipolar signature of the resonance in the amplitude of resulting current (read out as voltage). For each frequency value we measure the peak-to-peak amplitude of the signal five times. This value results in a precise measurement and more iterations add time to the acquisition without significantly improving the precision. The points and the uncertainties are the mean and the standard deviation of the different measurements, respectively. Figure~\ref{fig:model} shows a typical frequency scan observed in our setup. The bipolar resonance is clearly visible.  We fit Equation~\ref{eq:model} to this data to obtain $\omega_0$, from which we calculate the wire tension using Equation~\ref{eq:tension}. The parameters $c_1$, $c_2$ and $\Gamma$ are also allowed to vary in the fit.

\section{Implementation of the Setup}
\begin{figure}
  \centering    
  \includegraphics[width=0.45\textwidth]{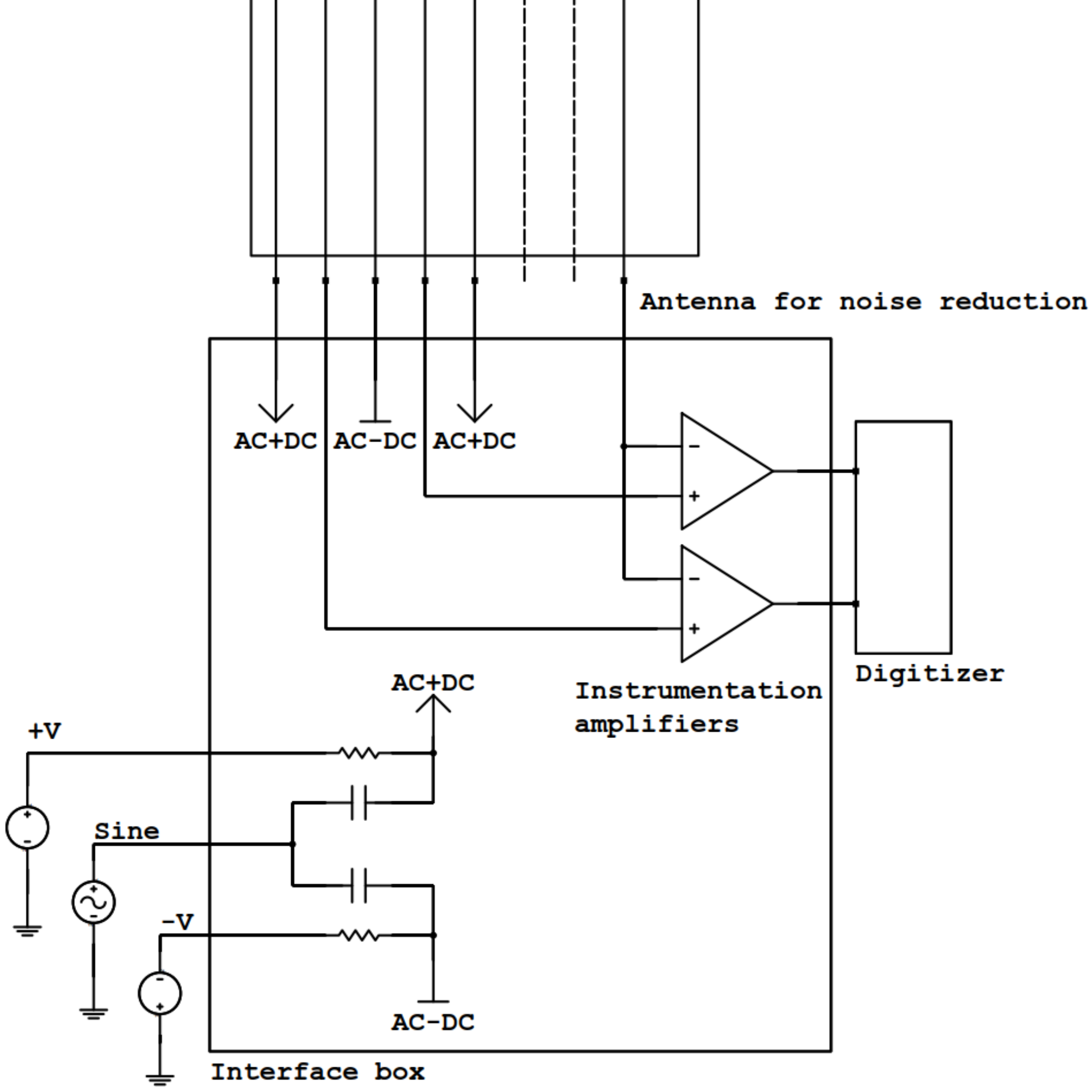}
  \caption{Simplified schematic of the measurement system.}
\label{fig:circuit}
\end{figure}

A schematic of the measurement system is shown in
Figure~\ref{fig:circuit}. The main components are:
\begin{itemize}
\item Two high voltage power supplies to provide the $\pm$ DC signal.
\item A standard sine wave generator with remote control interface (Keithley model 3390).
\item A linear AC high voltage amplifier to increase the sine wave amplitude above the 10\,V maximum value provided by the signal generator (Falco Systems model WMA-02).
\item A $24$--$30$~V DC power supply to power the AC amplifier ($< 150$~mA).
\item A bespoke interface box.
\item A DC power supply ($\pm 12$~V) to power the instrumentation amplifiers in the interface box.
\item A CAEN DT5740 Digitizer.
\item A PC running LabVIEW acquisition software.
\end{itemize}

The measurements described in this article are made using a custom stainless steel frame with anchored FR-4 boards at each end, which is similar to currently used wire plane designs, e.g., in the SBND detector~\cite{Antonello:2015lea}. Copper-beryllium wires with a diameter of 0.15\,mm are soldered to metallic pads at a 3\,mm pitch. The wires are connected to the electrical system through a bespoke interface box.

Within the interface box, the positive and negative DC voltages from the two high voltage power supplies are each combined with the amplified AC voltage from the signal generator to form the two bias voltages ($V_{\rm AC} + V_{\rm DC}$ and $V_{\rm AC}-V_{\rm DC}$). An extra wire is connected without bias as an antenna to subtract ambient noise from the measured voltages. The interface box connects the bias voltages to the adjacent wires and the wires under test to instrumentation amplifiers that buffer the signals and remove interference by subtracting the noise signal provided by the antenna wire.

The wires under test and the antenna wire are connected to the interface box via a cable harness comprising a screened ribbon cable to minimize coupling of the bias AC component between the wires. The bias voltages are connected using hook-up wire.
The buffered signals from the wires under test are then fed to the digitizer and onto the PC. A LabVIEW program running on the PC sweeps the injected AC signal through a defined range of frequencies and logs an average of the amplitude of the acquired signals for each frequency step.

One of the attractive features of this novel method is the option of measuring the tension of multiple wires at the same time. The number of wires that can be measured at once is only limited by the hardware cost per channel. The multi-channel system we have developed can read out 32 wires at a time whilst connected to 65 wires (every other wire has an AC$\pm$DC bias). This can easily be scaled up to any multiple of 32 (or 64 if using the VME version of the digitizer). With the described setup, our measurement time is about \unit[1]{s} for a scanning range of \unit[1]{Hz}.

\section{Performance of the system} \label{sec:performance}

\subsection{The signal features}
The novelty of the approach described here is the capability to measure the tension of multiple wires without ambiguity.  During a frequency scan, the fundamental frequency of each of the wires involved in the measurement should be observed through a resonant signal. If all the wires are at the same or very similar tensions, which is a typical feature in real-life detectors, the signal frequencies will lie very close together or overlap. The method using only two wires, proposed in Ref.~\cite{Ohama:1997ww}, makes it very complicated or even impossible to distinguish between the wires in the frequency scan as the resulting amplitudes are of comparable size.

\begin{figure}[h]
  \centering
\includegraphics[width=0.43\textwidth]{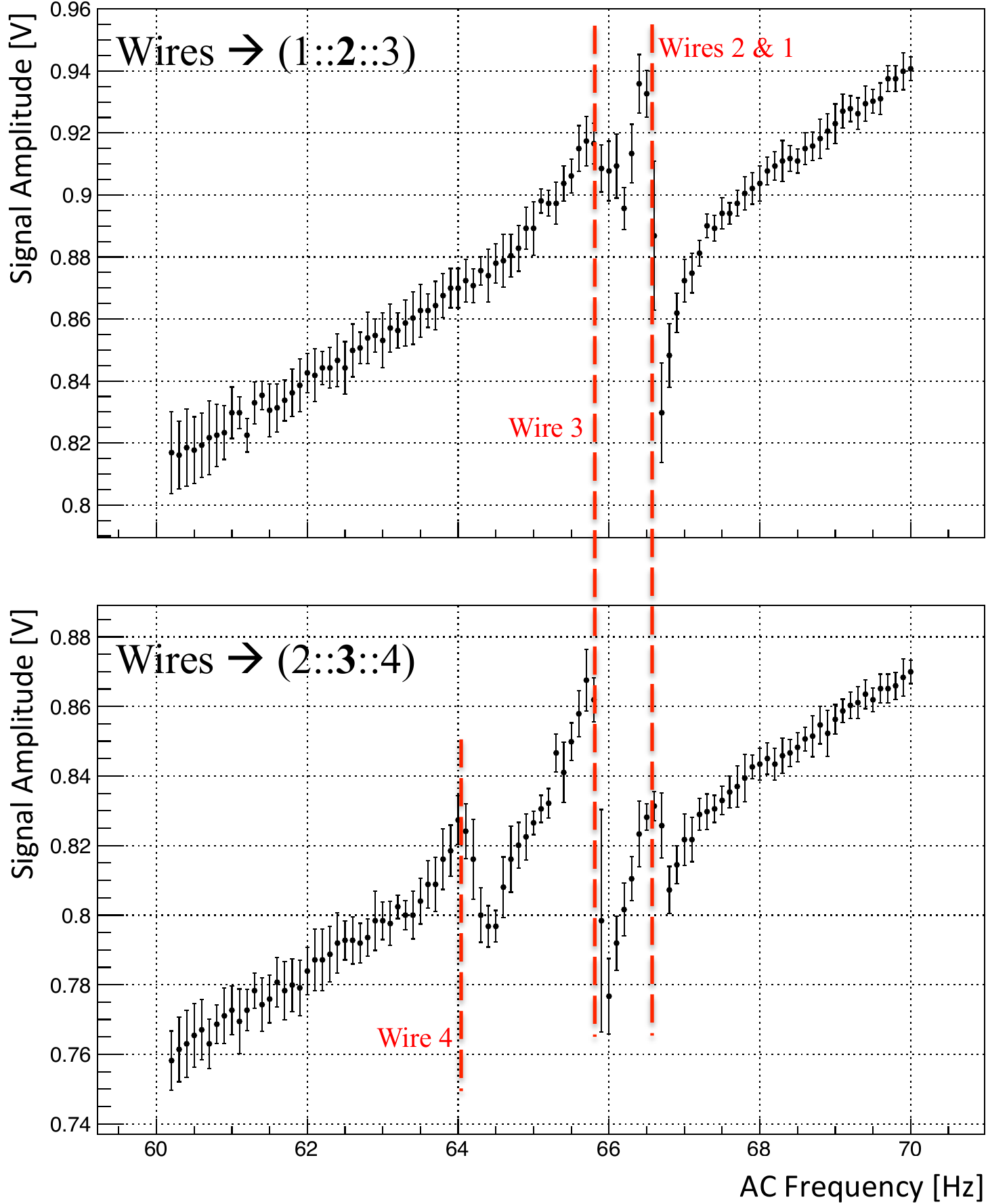}
  \caption{An illustration of the principle of decoupling the frequency of the wire under measurement from the biased wire frequencies. In the top panel wire 2 is measured and the large dip corresponds to its natural frequency. In the bottom panel wire 3 is being measured, making its dip significantly larger and the dip corresponding to wires 2 and 4 significantly smaller.}
\label{fig:signals}
\end{figure}

\begin{figure}[t]
\centering
\includegraphics[width=0.39\textwidth]{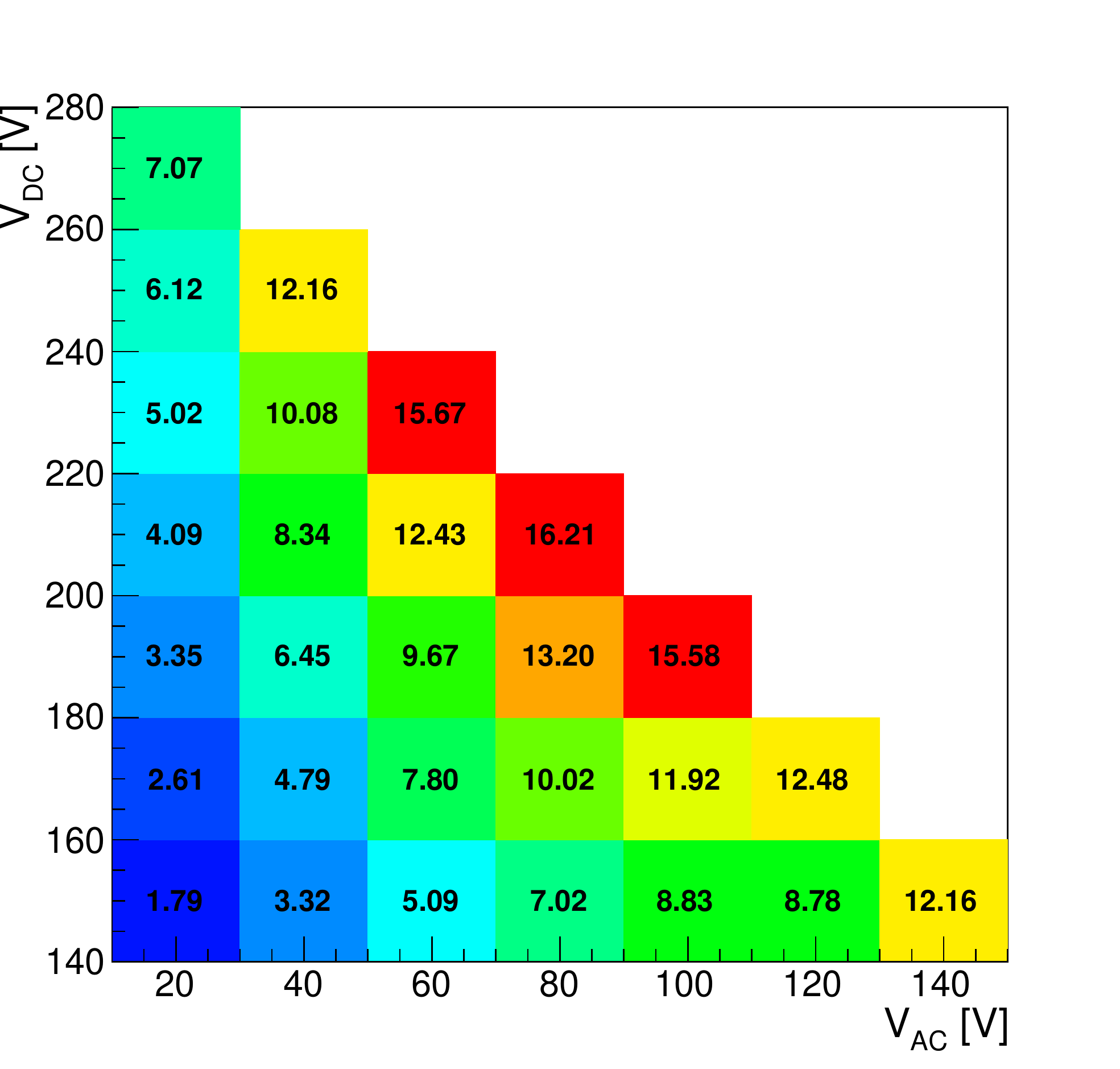}\\(a)\\
\includegraphics[width=0.39\textwidth]{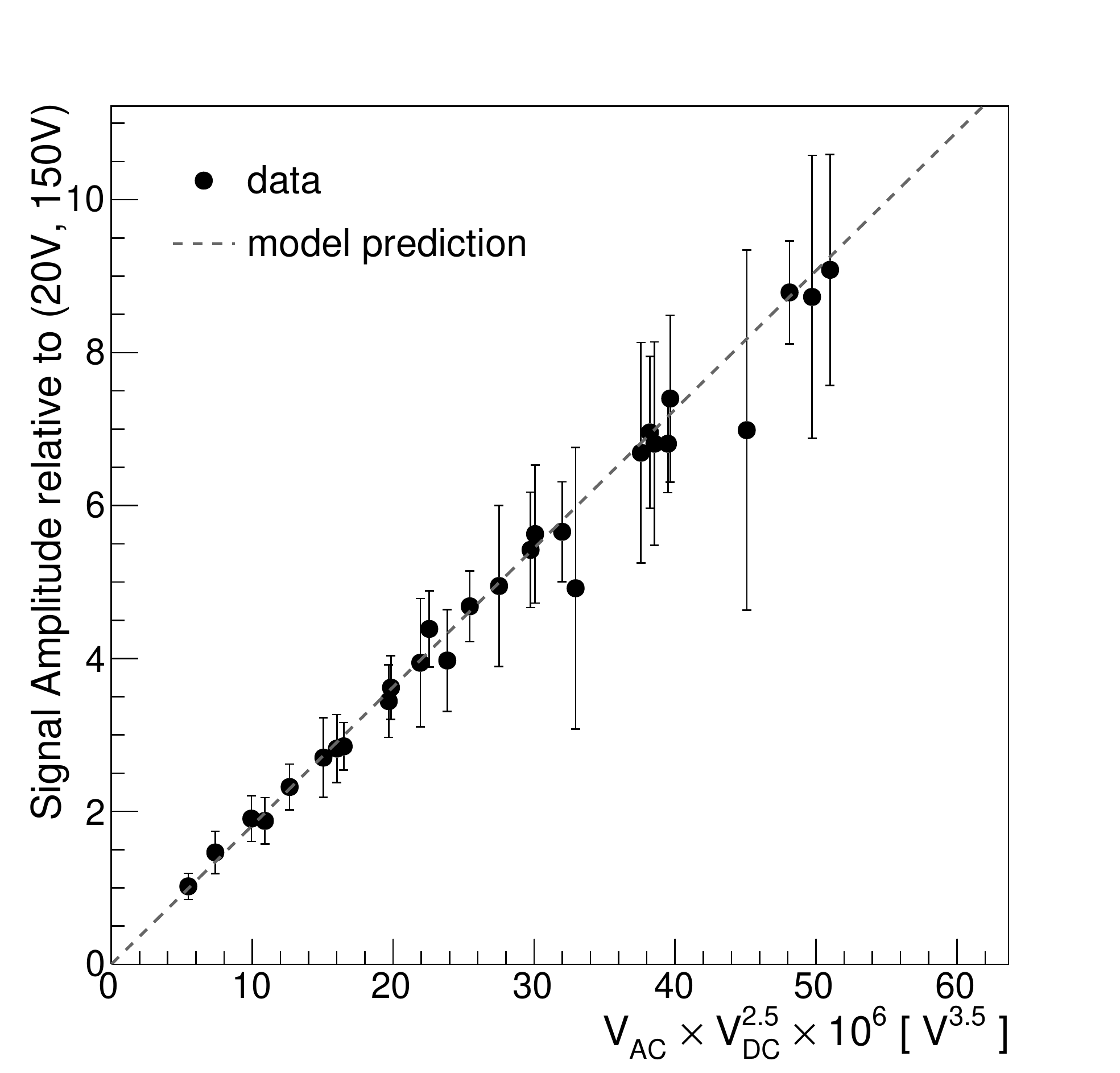}\\(b)\\
\caption{(a) Signal amplitudes (in mV) measured in bins of ($V_{\rm AC}, V_{\rm DC}$) and (b) as a function of $V_{\rm AC} \times V_{\rm DC}^{2.5}$. All measurements are quoted relative to the value in the bin (20\,V, 150\,V). Measurements are compared to the function defined by Equation~\ref{eq:model} (dashed line).}
\label{fig:validation}
\end{figure}

In the setup presented here, multiple wires are biased and grounded in parallel, see Figure~\ref{fig:circuit}, but each individual measurement uses three wires ($V_{\rm AC}+V_{\rm DC}$, ground, $V_{\rm AC}-V_{\rm DC}$), as illustrated in Figure~\ref{fig:cartoon}. The wire in the center of the three-wire configuration will produce a significantly higher signal than the other two, because of the opposite-sign electric fields applied on both sides of the wire. This feature makes possible to discriminate between the inner and outer wires and resolve degeneracies between the fundamental frequencies of adjacent wires. To illustrate this point, we built a system with four wires of 1.3\,m length each (typical length of wire segments in the ProtoDUNE single phase and SBND detectors). In order to demonstrate the central wire enhancement we prepared the wires with slightly varied tensions: (1) 5\,N, (2) 5\,N, (3) 4.8\,N, and (4) 4.6\,N. These tensions correspond to frequencies of 66.8\,Hz, 66.8\,Hz, 66\,Hz, and 64\,Hz, respectively which are visibly separate in a frequency scan. 

For this illustration we use relatively high voltages of 40\,V for the AC amplitude and $\pm 350$\,V for the DC voltage, to increase the amplitude of the signals. As a result the signal of the adjacent, non-grounded, wires is clearly visible. In the first case, shown in the top panel of Figure~\ref{fig:signals} we bias wire~(1) and wire~(3) at AC$\pm$DC and leave wire~(2) at ground. As a result we observe two distinct signals where the larger peak, at higher frequency and thus higher tension, is the combined signal of wires~(1) and (2), which are at the same tension (5\,N). The smaller peak is the signal of wire~(3). In the second case, shown in the bottom panel of Figure~\ref{fig:signals}, we instead bias wire~(2) and wire~(4) applying AC$\pm$DC voltages while leaving wire~(3) at ground, enhancing the signal from that wire. As expected, the large signal of wire~(3) in the bottom panel appears at the same frequency of the smaller signal in the top panel. In the bottom panel we observe two small peaks from the adjacent wire~(2) and wire~(4), as they have different tensions.  In both cases, the signal of the wire at ground is significantly larger than for the other wires. The disambiguation becomes even easier when using smaller voltages, because the signals of the adjacent stimulation wires become negligible leaving only the signal of the wire under evaluation. This can be observed in the signals shown in Figure~\ref{fig:model} and 
Figures~\ref{fig:wire2}--\ref{fig:segment}.

\subsection{Dependence on $V_{\rm AC}$ and $V_{\rm DC}$} \label{sec:voltage}
The $c_2$ term of the model described in Equation~\ref{eq:model}, derived assuming a simple two wire system, predicts that the amplitude of the output signal is proportional to the product of $V_{\rm AC}$ and $V_{\rm DC}^{2}$. To test if this assumption holds for our three wire system, we measure the amplitude in wires of 1\,m length under different bias conditions. We choose $V_{\rm AC} = 20\,V$ and $V_{\rm DC}=150\,V$ as a reference-point, which corresponds to an amplitude of about 1.8\,mV in our setup, and we increase the values of $V_{\rm AC}$ and $V_{\rm DC}$ in steps of 20\,V whilst keeping ($V_{\rm AC}+V_{\rm DC})<$~\unit[300]{V}. 

The results are shown in Figure~\ref{fig:validation}(a), where the signal amplitudes measured for different values of $V_{AC}$ and $V_{DC}$ are shown. This data set and the structure of Equation~\ref{eq:model} allow us to evaluate whether the signal amplitude depends on the product $V_{\rm AC} \times V_{\rm DC}^{2}$, as predicted by the two wire simplified model, or whether the power dependency of either term is different. To test this we considered the rows (constant $V_{\rm DC}$) and columns (constant $V_{\rm AC}$) of Figure~\ref{fig:validation}(a) as separate data sets and fit a power law behaviour to $V_{\rm AC}$ and $V_{\rm DC}$, respectively. We found a linear dependency on $V_{\rm AC}$ and that $V_{\rm DC}$ prefers a slightly steeper dependence than expected, namely $V_{\rm DC}^{2.5}$. This is demonstrated in Figure~\ref{fig:validation}(b) where for simplicity we have normalized the signal amplitudes to the value at the reference point (20\,V, 150\,V). The data show good agreement with the prediction of Equation~\ref{eq:model}, where we change the $c_2$ term to depend on $V_{\rm AC} \times V_{\rm DC}^{2.5}$. We can use this updated semi-empirical model to predict the signal size and to design a setup of the type proposed here given a particular performance requirement. This will be further explored in Section~\ref{sec:predictions}.

\subsection{Dependence on wire length} \label{sec:length}
\begin{figure}[htbp]
\centering
    \includegraphics[width=0.385\textwidth]{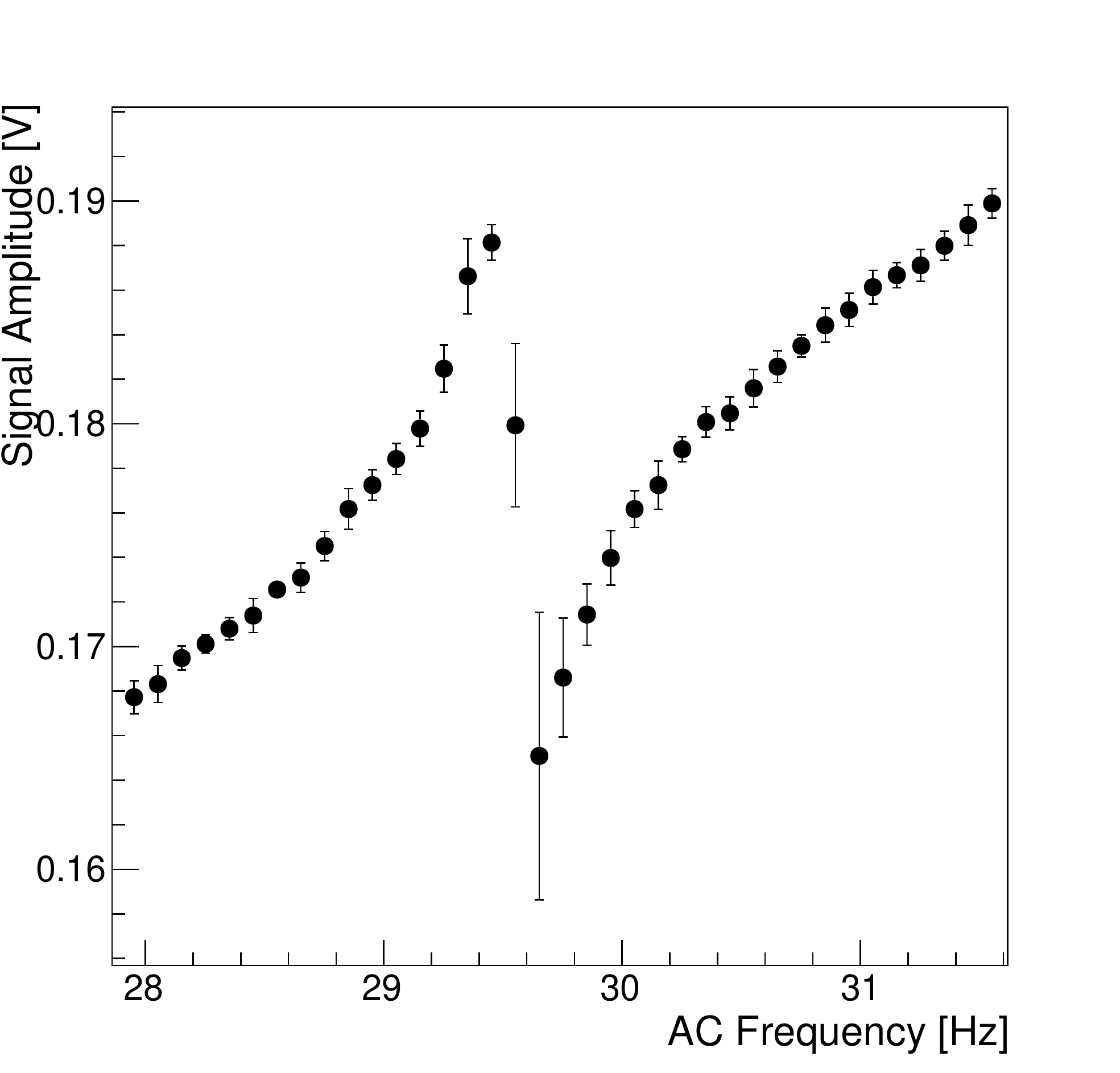}\\(a)\\
     \includegraphics[width=0.385\textwidth]{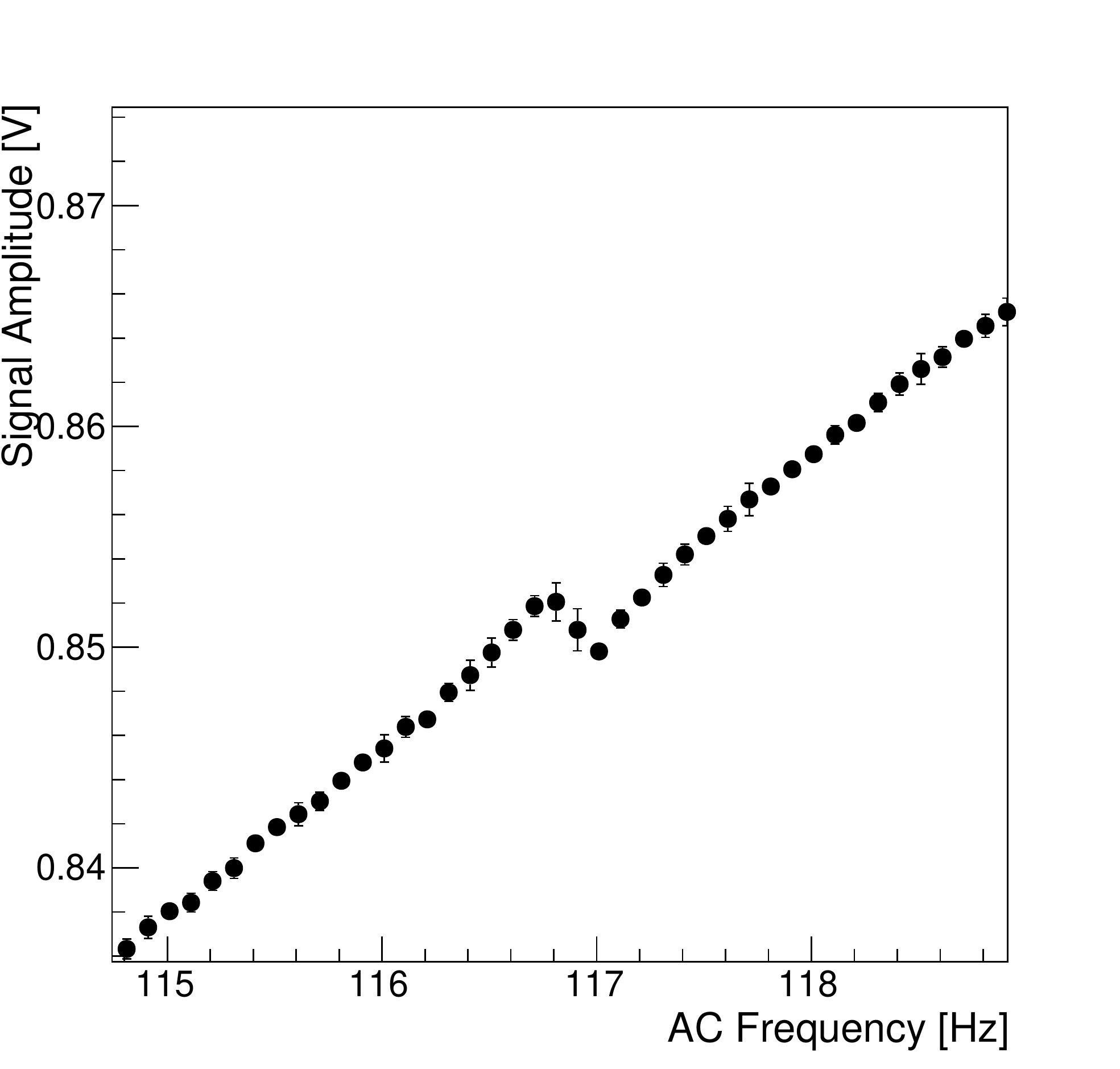}\\(b)\\
   \includegraphics[width=0.385\textwidth]{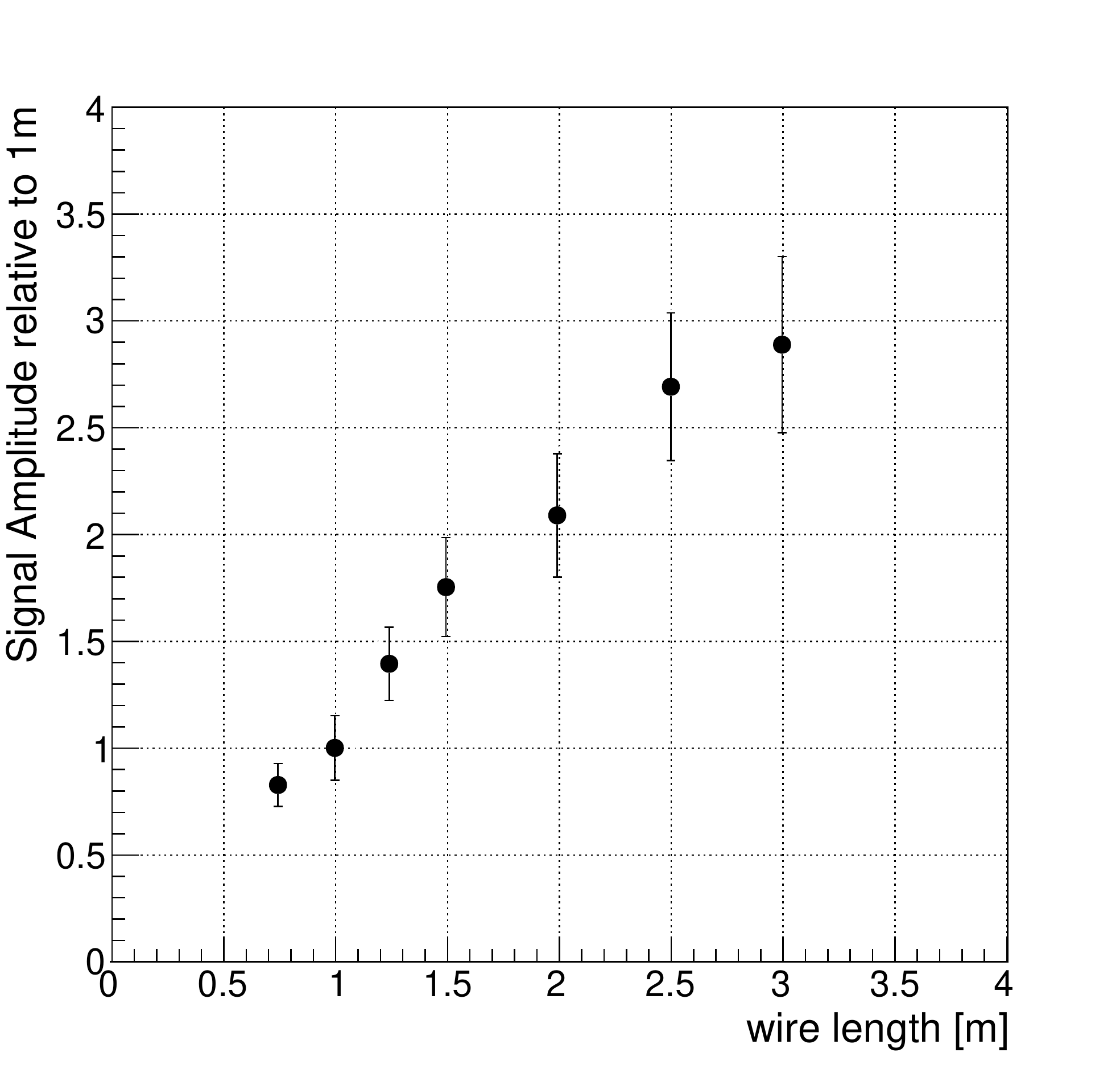}\\(c)\\
\caption{Examples of peak-to-peak amplitudes of signals as a function of frequency measured at a wire length of (a) 3\,m and (b) 0.75\,m. The wires are biased with $V_{\rm AC} \pm V_{\rm DC} = (40 \pm 150)$~V. (c) Measured signal amplitudes as a function of the wire lengths. Measurements are given relative to the value at 
a wire length of 1\,m.}
\label{fig:wire2}
\end{figure}

The signals shown in Figures~\ref{fig:wire2}(a) and (b) are the peak-to-peak voltage in the circuit at different frequencies for wires with a length of 3\,m and 0.75\,m, respectively, at voltages of $V_{\rm AC} =$\unit[40]{V} and $V_{\rm DC} =$\unit[150]{V}. As predicted by Equation~\ref{eq:tension}, the resonances occur at different frequencies, and produce a distinguishable signal in both cases. The signal amplitude is significantly larger for the longer wire. In Equation~\ref{eq:model}, the factors $c_2$, $\omega_0$ and $f(\omega)$ depend on the wire length. The dependency of the $f(\omega)$ factor is canceled by the dependency of $\omega_0$, leaving the $c_2$ term to govern the dependence of the resonance's amplitude on the wire length. The dependency of $c_2$ on the wire length arises due to the capacitance of the system: the capacitance between two adjacent wires is proportional to their lengths, hence we expect the dependence of the resonance amplitude on the wire length to be linear. 

To test this assumption, we study the effect of the wire length on the signal amplitude. Figure~\ref{fig:wire2}(c) shows the measured amplitudes for wires at the same tension and with lengths in the range of $0.75$--$3$\,m using the voltages $V_{\rm AC} = 40$\,V and $V_{\rm DC} = 150$~V. The amplitude is given relative to the operating point of $1$~m. We observe the expected linear relationship between amplitude and wire length, and we also see that measurements can be made for wires as short as 0.75\,m. In Section~\ref{sec:predictions}, we will extrapolate this dependence to determine the shortest wire length measurable at these voltages. 

\subsection{Multi-segment wires}
The design of large-scale TPCs requires long wires with lengths of $5$~m in the MicroBooNE detector~\cite{Acciarri:2016smi} or $6$~m in the protoDUNE single-phase detector~\cite{Abi:2017aow}. Wires of this length could be prone to mechanical deflection due to gravity, electrostatic forces, or even movement induced by argon flow. To reduce such effects, a support structure is introduced at regular intervals along the length of the wire plane. An example of such a structure is a series of plastic combs mounted on cross-braces that divide the long wires into segments. This minimizes any sagging and, should a wire break, mitigates the impact of broken wires by restricting the impact to a smaller region.

\begin{figure}
  \centering
    \includegraphics[width=0.44\textwidth]{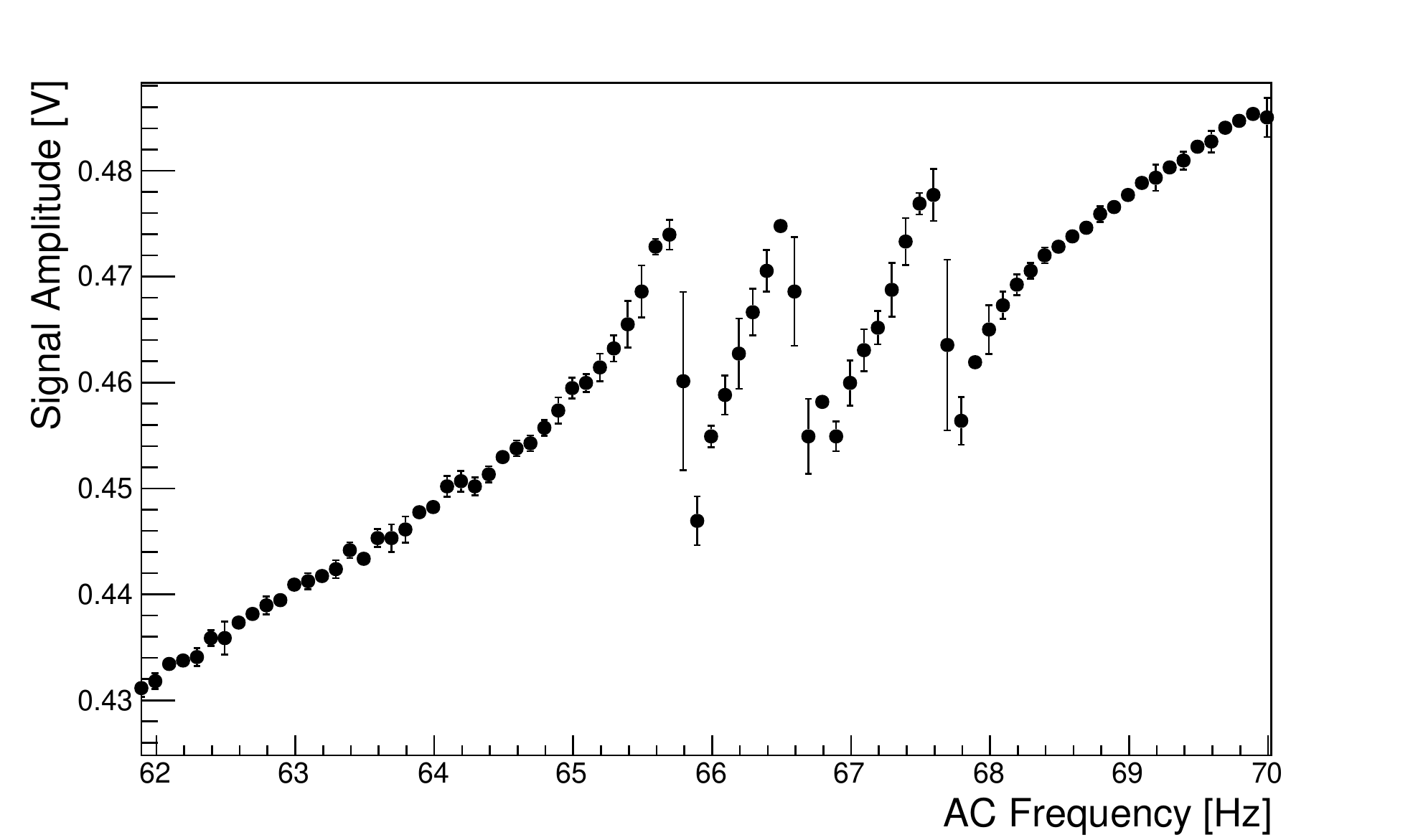}
  \caption{Signals in wires that have been split up into three segments of about 1.3\,m length each. The resonant signal from each segment is clearly visible. The wires were biased with $V_{\rm AC} = 40$\,V and $V_{\rm DC} = 250$\,V.}
\label{fig:segment}
\end{figure}

The effect of such combs on the tension measurements is that each wire is split into multiple shorter wires with lengths determined by the spacings between the combs. These wires can be at different tensions if they are glued to the combs. A setup using the laser method would need to measure each segment separately. The electrical  measurement method is sensitive to the fundamental frequency of all the individual segments connected electrically. We have tested this effect using a setup where wires of 4\,m length are segmented into three units of about 1.3\,m using dielectric combs. Figure~\ref{fig:segment} shows the signal obtained for a segmented wire with well separated resonance peaks for each segment. This demonstrates that the electrical method can be used to measure the tension of all segments of wires divided by combs with a single measurement.

\subsection{Resolution of the method}
\begin{figure}
\centering
\includegraphics[width=0.39\textwidth]{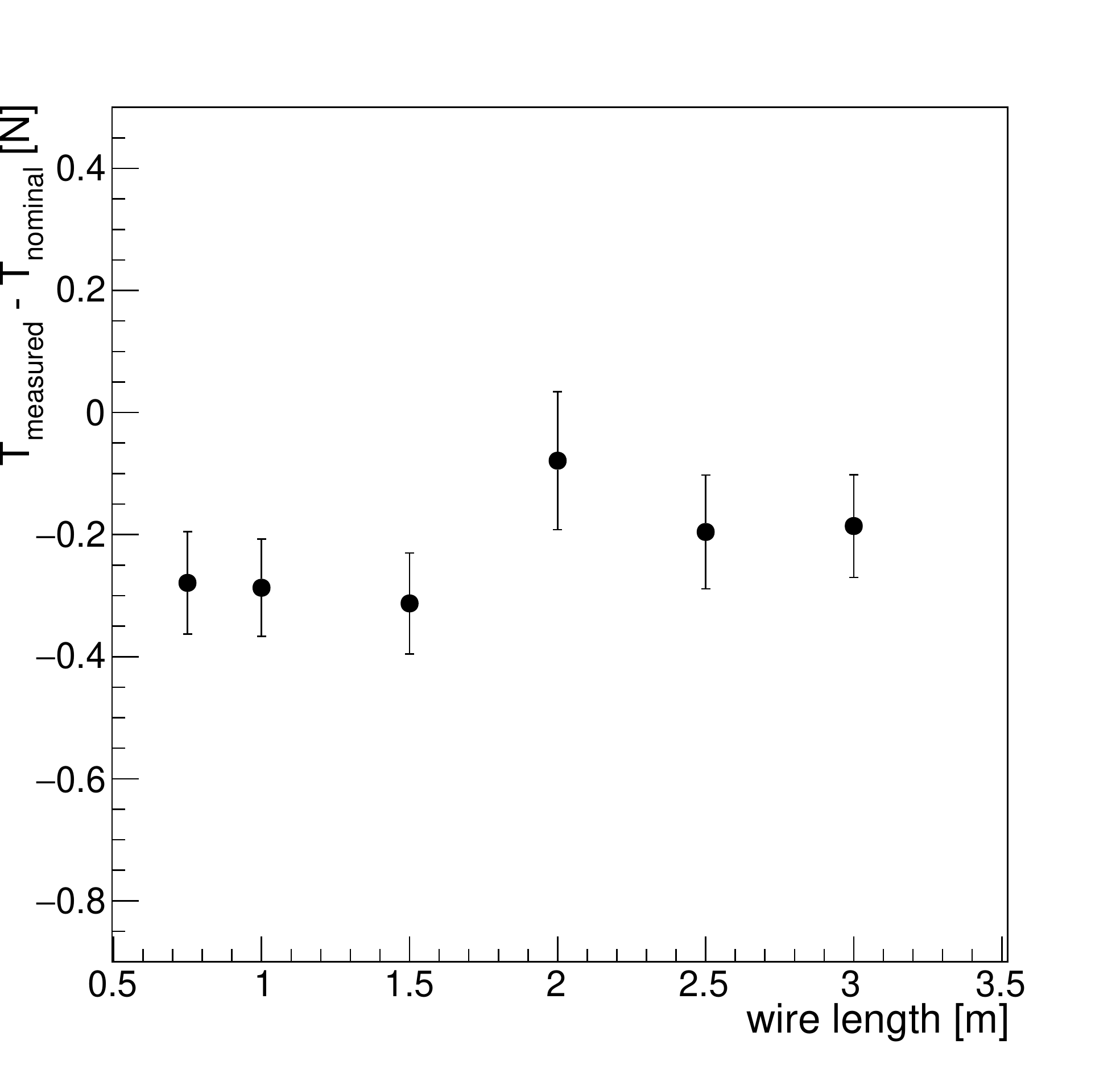}\\(a)\\
\includegraphics[width=0.39\textwidth]{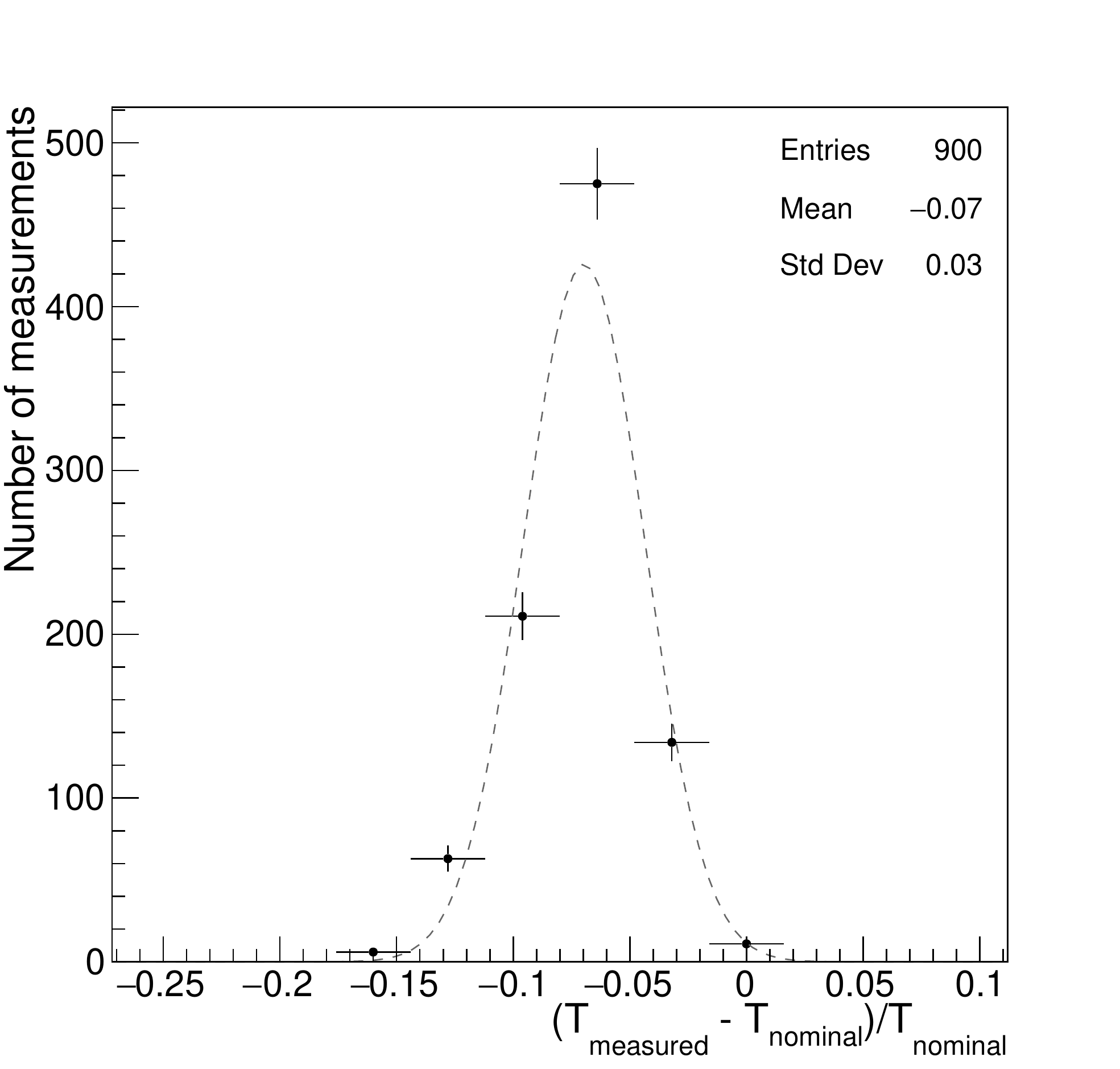}\\(b)\\
\caption{(a) Difference between the measured and nominal tension for different wire lengths. 
(b) Fit of a Gaussian function to the relative tension bias for wire lengths between 0.75\,m and 3\,m.}
\label{fig:resolution}
\end{figure}

We determine the resolution of our method for different wire lengths by repeatedly measuring the tensions of a sample of ten wires of the same length. The tension is chosen to be 5\,N as this value is typical for the nominal tension of wires in LArTPCs~\cite{Acciarri:2016smi,Abi:2017aow}. The tension of the wires is set by soldering the wire to the wire-bonding-board at one end whilst it is strung with weights attached to the other end. We assume the uncertainty on such setting of the tension to be negligible. The wires are biased with $V_{\rm AC} = 40$~V and $V_{\rm DC}=150$~V. We repeat this measurement at different lengths of the sets of wires: 0.75\,m, 1\,m, 1.5\,m, 2\,m, 2.5\,m, and 3\,m. The results are shown in Figure~\ref{fig:resolution}. We observe a small reconstruction bias that depends weakly on wire length. The method systematically underestimates wire tensions by $0.2$--$0.3$\,N, or $7\%$. A Gaussian fit to the measurements, shown in Figure~\ref{fig:resolution}b,  yields a spread of the measurements, and therefore a resolution of the technique, of $\approx 3\%$.

\subsection{Limits of Applicability} \label{sec:predictions}

The results from Sections~\ref{sec:voltage} and \ref{sec:length} allow us to extrapolate the model to find the limits of the applicability of our system. We determine the lowest combination of AC and DC voltages and the shortest wire length at which the electrical method can be applied. To ensure the validity of the method, we require that $95\%$ of all measurements return a tension measurement within $5\%$ of the true tension. The success rate of such a measurement would depend on the amplitude of the signal and the noise.  

We use a toy Monte Carlo code to generate signals as would be observed by the electronic setup. The amplitude is calculated using the dependencies presented in Figures~\ref{fig:validation} and \ref{fig:wire2}. The noise is modeled using measurements in our system. We find that the noise depends weakly on the AC amplitude as RMS/mV$ = (0.45 + 0.0025 V_{\rm AC}$/V). The dependence on frequency and wire length can be neglected. We shift the position of the signal resonance peak in the simulation within the measurement interval of \unit[0.1]{Hz} by a random phase.

To determine the shortest wire length, we chose the operating voltages of $V_{\rm AC} = 40$\,V and $V_{\rm DC} = 150$\,V as in Section~\ref{sec:length}. Using the model we determine that the method should be applicable down to a wire length of 50\,cm. Measuring shorter wire lengths is possible, but would require applying higher bias voltages. We use wires of 1\,m length and the dependence of the amplitude on voltage observed in Section~\ref{sec:voltage} to find that the minimum values of $V_{\rm AC} = 20$\,V and $V_{\rm DC} = 150$\,V are close to the sensitivity limit of the technique, corresponding to a signal amplitude of about $1.8$\,mV. The electrical setup presented here is tested with a wire pitch of $3$~mm, a decrease of the signal amplitude is expected for a larger wire pitch.

\section{Cryogenic Operation} \label{sec:cold}

The electrical technique can be applied within a closed cryostat and at cryogenic temperatures, which
is a novel feature and important advantage. The possibility of performing a measurement under such conditions, e.g., during the cool down of the liquid argon allows us to track the tension in real time. If the thermal contraction of the different materials significantly alters the tension of the wires, risking the integrity of the system and potentially modifying the performance of a detector, we could perform an intervention to mitigate the effect. 

To demonstrate this capability, we tested our setup down to a temperature comparable with that of liquid argon ($-186^\circ$\,C) by creating a stable environment of cooled down gaseous nitrogen. A sample of CuBe wires are soldered to FR-4 boards anchored at the two ends of a stainless steel frame defining a length of 72\,cm for the wires. The coefficients of thermal expansion of CuBe alloy, FR-4 and stainless steel are similar. The materials in this setup were chosen to minimize the potential wire tension increment during the cool-down. The system is immersed inside a vacuum-jacketed dewar which is filled with a cold gas atmosphere. Liquid nitrogen was added to the system in controlled amounts to slowly decrease the temperature. The dewar is equipped with four Pt1000 temperature sensors uniformly distributed between the bottom and the top of the steel frame to monitor the temperature along the full length of the wires. During the cool down we maintain a gaseous atmosphere, keeping the liquid at all times at a level below the wires, to avoid additional damping from the viscosity of the liquid. We perform tension measurements in regions of relative equilibrium where the temperature difference between the lowest and highest Pt1000 sensors is less then 30$^\circ$\,C.

\begin{figure}[htbp]
  \centering
    \includegraphics[width=0.45\textwidth]{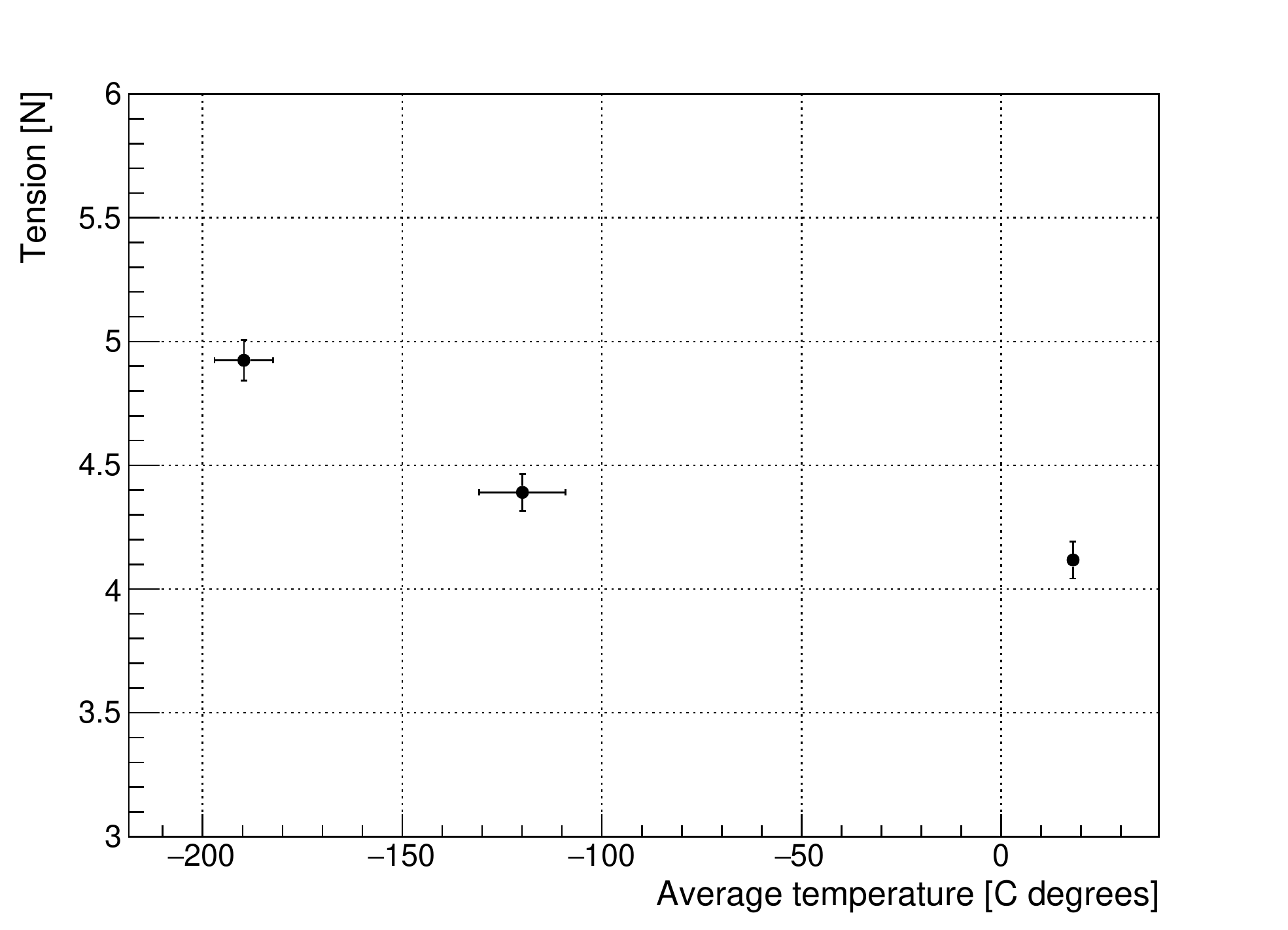}
  \caption{Wire tension values as function of temperature measured during the cold test.}
\label{fig:cold}
\end{figure}

Figure~\ref{fig:cold} shows the first results of the tension measurements at cryogenic temperatures. The data points and error bars represent the mean and standard deviations of our measurements. The equilibrium condition is satisfied at around $-120^{\circ}$\,C and $-190^{\circ}$\,C. At these temperatures the tension of the wires, compared to room temperature, increases by about $6\%$ and $20\%$, respectively. This is a larger increase than expected from the change in length of the wire caused by the thermal contraction derived using Equation~\ref{eq:tension}.

\section{Conclusions}

We have designed and developed an electrical method to precisely measure the tensions of multiple wires simultaneously with a resolution of about $3\%$ applying voltages of a few hundred volts. The method uses a frequency scan of biased wires of alternating potential and allows a measurement to be completed within minutes, significantly reducing cost and time compared to wire-by-wire measurements using lasers. The method also removes the need for a physical disturbance of the wires. These features make this procedure advantageous for the measurements of tensions of wires in large wire chambers, with thousands of wires. We study the behaviour of our system with respect to the bias voltages applied and the wire length and we develop a model that allows us to predict the amplitude of the signal.  We show that a measurement can be made down to wires as short as 50\,cm. Finally, we demonstrate that this technique can be used to measure wire tension at cryogenic temperatures, which has not been feasible before and should be applicable during cool down of large liquid-argon TPCs. 

\section*{Acknowledgements}

This work was in part funded by the Royal Society and the Science and Technology Facilities Council (STFC). V.B. is funded by a Presidential Doctoral Scholarship at the University of Manchester. We thank Claire Fuzipeg for the development of the LabVIEW DAQ software. We thank Dr. Roxanne Guenette for her insightful comments on the manuscript. 

\bibliographystyle{unsrt}
\bibliography{bibliography}

\begin{thebibliography}{10}

\bibitem{Rubbia:1977zz}
C.~Rubbia.
\newblock {The Liquid Argon Time Projection Chamber: A New Concept for Neutrino
  Detectors}.
\newblock 1977.

\bibitem{Antonello:2015lea}
M.~Antonello et~al.
\newblock {A Proposal for a Three Detector Short-Baseline Neutrino Oscillation
  Program in the Fermilab Booster Neutrino Beam}.
\newblock 2015.

\bibitem{Acciarri:2016ooe}
R.~Acciarri et~al.
\newblock {Long-Baseline Neutrino Facility (LBNF) and Deep Underground Neutrino
  Experiment (DUNE)}.
\newblock 2016.

\bibitem{Acciarri:2017sde}
R.~Acciarri et~al.
\newblock {Noise Characterization and Filtering in the MicroBooNE Liquid Argon
  TPC}.
\newblock {\em JINST}, 12(08):P08003, 2017.

\bibitem{Acciarri:2016ugk}
R.~Acciarri et~al.
\newblock {Construction and Assembly of the Wire Planes for the MicroBooNE Time
  Projection Chamber}.
\newblock {\em JINST}, 12(03):T03003, 2017.

\bibitem{Baldini:2007iba}
W.~Baldini et~al.
\newblock {A Laser Based Instrument for MWPC Wire Tension Measurement}.
\newblock 2007.

\bibitem{Wang:2017mbk}
Xu~Wang, Fuwang Shen, Shuai Wang, Cunfeng Feng, Changyu Li, Peng Lu, Jim
  Thomas, Qinghua Xu, and Chengguang Zhu.
\newblock {Design and implementation of wire tension measurement system for
  MWPCs used in the STAR iTPC upgrade}.
\newblock {\em Nucl. Instrum. Meth.}, A859:90--94, 2017.

\bibitem{Ohama:1997ww}
T.~Ohama, N.~Ishihara, S.~Takeda, H.~Okuma, and K.~Konno.
\newblock {Electrical measurement of wire tension in a multiwire drift
  chamber}.
\newblock {\em Nucl. Instrum. Meth.}, A410:175--178, 1998.

\bibitem{Hoshi:1985}
Minoru~SATOH Yoshimoto~HOSHI and Masato HIGUCHI.
\newblock {A simple method for measuring wire tension in drift tubes}.
\newblock {\em Nucl. Instrum. Meth.}, A236:82--84, 1985.

\bibitem{Brinkley:1996xb}
B.~Brinkley, J.~Busenitz, and G.~Zilizi.
\newblock {Wire tension measurement using voltage switching}.
\newblock {\em Nucl. Instrum. Meth.}, A373:23--29, 1996.

\bibitem{Lang:1998yw}
K.~Lang, J.~Ting, and V.~Vassilakopoulos.
\newblock {A Technique of direct tension measurement of a strung fine wire}.
\newblock {\em Nucl. Instrum. Meth.}, A420:392--401, 1999.

\bibitem{Andryakov:1998vk}
A.~Andryakov et~al.
\newblock {Electrostatic digital method of wire tension measurement for KLOE
  drift chamber}.
\newblock {\em Nucl. Instrum. Meth.}, A409:63--64, 1998.

\bibitem{Acciarri:2016smi}
R.~Acciarri et~al.
\newblock {Design and Construction of the MicroBooNE Detector}.
\newblock {\em JINST}, 12(02):P02017, 2017.

\bibitem{Abi:2017aow}
B.~Abi et~al.
\newblock {The Single-Phase ProtoDUNE Technical Design Report}.
\newblock 2017.

\end{thebibliography}

\end{document}